\documentclass[12pt]{article}

\begin{document}

\title{\bf
Using virtual processors for SPMD parallel programs}

\date{December 2003}

\author{\bf Gianluca Argentini \\
\normalsize gianluca.argentini@riellogroup.com \\
\textit{New Technologies and Models}\\
Information \& Communication Technology Department\\
\textit{Riello Group}, 35044 Legnago (Verona), Italy}

\maketitle

\begin{abstract}
In this paper I describe some results on the use of virtual processors technology 
for parallelize some SPMD computational programs. The tested technology is the INTEL
Hyper Threading on real processors, and the programs are MATLAB scripts for floating
points computation. The conclusions of the work concern on the utility and limits of
the used approach. The main result is that using virtual processors is a good technique
for improving parallel programs not only for memory-based computations, but in the case 
of massive disk-storage operations too.
\end{abstract}

\section{Introduction}
The processors virtualization technology permits to split a real physical processor
into two virtual chips, so that the operating system, as MS Windows or Linux, of a 
computer can use the virtual processors as two real chips. Example of such technology
is Intel's Hyper Threading [1]. The hardware can so be considered as a symmetric 
multi-processor machine and the software can use it as a true parallel 
environment.

  In this work I show some results obtained with parallel computations using 
Matlab [2] programs on Intel technology. The physical and logical characteristics of 
the used machine are presented in the following tables:\\

\begin{tabular}{|ll|}
\hline
\textbf{Hardware}		& \\
\hline
Type				&	HP Compaq ProLiant DL380 \\
Processors	&	2 Intel Xeon 2.40 GHz \\
Ram					&	1.5 GB \\
Storage			&	6 SCSI disks 36.5 GB - Raid 5 \\
\hline
\end{tabular} \\ \\

\begin{tabular}{|ll|}
\hline
\textbf{Software}   & \\
\hline
Operating System		& MS Windows Server 2000 \\
Matlab							& v. 5.3 \\
\hline
\end{tabular} \\ \\

  The Matlab programs used for these experiments was based on cycles of floating-point 
computations.

\section{The parallel Matlab environment}
The package Matlab has not a native support for parallel elaboration and
multithreading [3]. Yet, there are some extensions, as tools and libraries [4], for
the use of a parallel environment on multi-processors hardware. With the biprocessor
machine I have preferred the method of splitting a given computation on multiple instances
of the runtime Matlab program. A single master instance starts the slave copies and 
assigns to each of them the same set of instructions on different sets of data. Hence I
have simulated a SPMD computation on the machine. 

  In this way the parallel environment is simple, because there is not need of external 
libraries or calls to interfaces, and flexible, because to a single slave copy it can be 
assigned a set of different instructions for realizing a MPMD computation.

  With this method the exchange of messages among independent processes is a problem. 
The only way to communicate from one Matlab copy to another is the use of shared files. 
In a second type of experiments I show that this method is not critical for the 
time execution if one uses fast mass-storage as SCSI or FiberChannel systems.

\subsection{The SPMD programs}
In the experiments I have defined a \textit{master} Matlab function which writes 
to a shared file system the .m scripts to be executed by \textit{slaves} Matlab copies.
These copies are launched in background mode for the parallel execution. The master 
program controls the end of the computations using a simple set of lock-files. 
The slaves finish their work, save on files the results and cancel the own lock-file. 
The master reads the sets of data from these files for other possible computations. 
Now I describe the principal code of the program.\\
  
  This is the declaration of the function \textit{masterf}. The \textit{lockarray} 
variable is an array for testing the presence of the lock-files during the slaves 
computation. The \textit{finalres} is an array for the collection of the partial
results from slaves. The string \textit{computing} is the mathematical expression
to use in the computation.

\scriptsize
\textbf{\\
function [elapsedtime,totaltime,executiontime]=masterf(nproc,maxvalue,step,computing)\\
\%\\
\% MASTERF: master function for parallel background computation.\\
\%\\
\% sintax:\\
\% [elapsedtime,totaltime,executiontime]=masterf(nproc,maxvalue,step,computing)\\
\%\\ 
\% input parameters:\\
\%\\
\% nproc = number of processes;\\
\% maxvalue = sup-limitation of the data-array to process; the inf-limitation is 0;\\
\% step = difference from two consecutive numbers in the data-array;\\
\% computing = the string of the mathematical expression to compute;\\
\%\\
\% output parameters:\\
\%\\
\% elapsedtime = total elapsed time to complete the execution of the computation;\\
\% totaltime = sum of the single slaves CPU-time to complete the single computation;\\
\% executiontime = single slaves CPU-time to complete the assigned computation;\\
\\
ostype=computer;\\
tottime=0.;\\
lockarray=0:nproc-1;\\
numbervalues=maxvalue;\\
computingstring=[' ' computing];\\
finalres=[];}\\

\normalsize
After the assignment of the own value to variable \textit{workdir}, working directory 
of Matlab, a cycle writes on storage the slaves lock-files.

\scriptsize
\textbf{\\
for i=0:nproc-1\\
\indent filelock = strcat(workdir,'filelock',int2str(i));\\
\indent fid=fopen(filelock,'wr');\\
\indent fwrite(fid,'');\\
\indent fclose(fid);\\
end}\\

\normalsize
In the next fragment of program, the master sets the commands for the writing 
of an appropriate Matlab .m script for every slave process. Such script contains
the instruction for determining the CPU-time spent on calculus, the expression 
of the mathematical computation, the instruction to save on storage the
data computed and the CPU-time, finally the instruction to delete the lock-file.

\scriptsize
\textbf{\\
for i=0:nproc-1\\
\indent  if (i==0) middlestep=0; else middlestep=1; end\\
\indent  infdata=i*(numbervalues/nproc) + middlestep*step;\\
\indent  supdata=(i+1)*(numbervalues/nproc);\\
\indent  fileworker = strcat(workdir,'fileworker',int2str(i),'.m');\\
\indent	 commandworkertmp = ...\\
\indent \indent      strcat('x=',num2str(infdata),':',num2str(step),':',num2str(supdata),...\\
\indent \indent      '; t1=cputime; ',computingstring,...\\
\indent \indent      '; t2=cputime-t1; save out',int2str(i));\\
\indent  commandworker = ['cd ' workdir '; ' commandworkertmp ...\\
\indent \indent      ' y t2; ' 'delete filelock'int2str(i) '; exit;'];\\
\indent  fid = fopen(fileworker,'wt');\\
\indent  fwrite(fid,commandworker);\\
\indent	 fclose(fid);\\
end}\\

\normalsize
After the instructions for determining the CPU-time and the elapsed-time
(\textit{tic}) spent by the master program, a cycle launches the slaves Matlab runtimes.
In the case of Windows operating system, the \textit{startcommand} string is "start", an
OS command for the background running of an executable program, and the \textit{osstring}
string is "dos". In the case of Unix-like operating system, the string are "sh" and
"unix" respectively. Each slave executes immediately the \textit{fileworker} script, 
as shown by the Matlab "-r" parameter.

\scriptsize
\textbf{\\
t1 = cputime;\\
tic;\\
for i=0:nproc-1\\
\indent   fileworker = strcat('fileworker',int2str(i));\\
\indent   commandrun = [startcommand ' matlab -minimize -r ' fileworker];\\
\indent   eval(strcat([osstring,'(','''',commandrun,'''',');']));\\
end}\\

\normalsize
In the next fragment of code the master program executes a cycle for determining
the end of slaves computations. It controls if the \textit{lockarray} variable has
some process's rank non negative. In this case, it attempts to open the relative 
lock-file; if the file still exists, the master closes it, else the lockarray 
process position is set to -1. The \textit{pause} instruction can be useful for
avoiding an excessive frequency, hence an high cpu-time consuming, in the "while"
cycle. 

\scriptsize
\textbf{\\
lockarraytmp=find(lockarray \begin{math}>\end{math} -1);\\
while (length(lockarraytmp) \begin{math}>\end{math} 0)\\
\indent   pause(.1);\\
\indent   for i=lockarraytmp\\		
\indent \indent      fid = fopen(strcat('filelock',int2str(i-1)),'r');\\
\indent \indent      if (fid \begin{math}<\end{math} 0)\\ 
\indent \indent \indent         lockarray(i) = -1;\\
\indent \indent      else\\
\indent \indent \indent       	fclose(fid);\\
\indent \indent      end\\
\indent  end\\
\indent  lockarraytmp=find(lockarray \begin{math}>\end{math} -1);\\
end}\\

\normalsize
At the end, the master reads the partial slaves computation outputs and stores 
them in an array.
At this point the master cpu-time and elapsed time are registered too. The total 
execution time is defined as sum of the slaves computation cpu-time, and is useful for
comparison with the execution time in the case \begin{math}\textit{nproc} = 1\end{math}.
The single slave execution time is defined as the arithmetic mean of all the partial
execution times.

\scriptsize
\textbf{\\
for i=0:nproc-1\\
\indent   partialres = load(strcat('out',int2str(i)));\\
\indent   finalres = [finalres partialres];\\
end\\
\\
elapsedtime = toc;\\
totaltime = cputime - t1;\\
\\
for i=0:nproc-1\\
\indent   fps = load(strcat('out',int2str(i)));\\
\indent   tottime = tottime + fps.t2;\\
\indent   executiontime = tottime/nproc;\\
end}\\

\normalsize

\section{Tests and results}
For the tests I have used the following values for the \textit{masterf} parameters:\\
\\
\textit{nproc}: from 1 to 8;\\
\textit{maxvalue}: m * 10000, where m = 1, 2, 3;\\
\textit{step}: 0.001;\\
\textit{computing}: \begin{math} y = 5432.060708*\cos((\sin(x^{9.876}))^{-1.2345}) \end{math}.\\

I have also tested the program without the slaves saving of partial computations results
and their final master load, for determining the influence of the I/O storage operations
on the times of execution.\\

The following are the results obtained with four tests for every type of experiment. The values
are arithmetic means approximated to two decimals and they are expressed in seconds; 
the numbers from 1 to 8 are the value of \textit{nproc}, while "m" is the factor parameter
in the \textit{maxvalue} expression.\\
I have not reported the elapsed-times, because they weren't different from the cpu-times
registered, probably due to the fact that, during the experiments, the server was dedicated
only to the computations.\\

\scriptsize
\textbf{Tables of results.\\
\indent All the values are expressed in seconds.}\\
\normalsize

\textbf{1.a} Medium execution cpu-times for process, no data storage:\\

\begin{tabular}{|l|llllllll|}
\hline
\textbf{m} & \textbf{1} & \textbf{2} & \textbf{3} & \textbf{4} & \textbf{5}
& \textbf{6} & \textbf{7} & \textbf{8}\\
\hline
\scriptsize{1} & \scriptsize{39.32} & \scriptsize{20.89} & \scriptsize{14.60} & 
\scriptsize{11.49} & \scriptsize{11.86} & \scriptsize{8.73} & \scriptsize{9.15} &
\scriptsize{7.29}\\
\hline
\scriptsize{2} & \scriptsize{77.56} & \scriptsize{40.83} & \scriptsize{29.20} & 
\scriptsize{23.41} & \scriptsize{22.31} & \scriptsize{23.19} & \scriptsize{17.93} &
\scriptsize{19.02}\\
\hline
\scriptsize{3} & \scriptsize{137.75} & \scriptsize{69.70} & \scriptsize{51.67} & 
\scriptsize{35.28} & \scriptsize{34.91} & \scriptsize{36.98} & \scriptsize{32.46} &
\scriptsize{34.86}\\
\hline
\end{tabular} \\ \\

\textbf{2.a} Total execution cpu-times, no data storage:\\

\begin{tabular}{|l|llllllll|}
\hline
\textbf{m} & \textbf{1} & \textbf{2} & \textbf{3} & \textbf{4} & \textbf{5}
& \textbf{6} & \textbf{7} & \textbf{8}\\
\hline
\scriptsize{1} & \scriptsize{41.01} & \scriptsize{22.11} & \scriptsize{16.19} & 
\scriptsize{13.67} & \scriptsize{18.42} & \scriptsize{17.10} & \scriptsize{18.17} &
\scriptsize{16.20}\\
\hline
\scriptsize{2} & \scriptsize{78.40} & \scriptsize{41.89} & \scriptsize{30.74} & 
\scriptsize{24.81} & \scriptsize{26.78} & \scriptsize{34.69} & \scriptsize{33.68} &
\scriptsize{32.95}\\
\hline
\scriptsize{3} & \scriptsize{139.05} & \scriptsize{75.66} & \scriptsize{59.38} & 
\scriptsize{38.83} & \scriptsize{40.28} & \scriptsize{49.69} & \scriptsize{48.98} &
\scriptsize{48.27}\\
\hline
\end{tabular} \\ \\

\textbf{1.b} Medium execution cpu-times for process, with data storage:\\

\begin{tabular}{|l|llllllll|}
\hline
\textbf{m} & \textbf{1} & \textbf{2} & \textbf{3} & \textbf{4} & \textbf{5}
& \textbf{6} & \textbf{7} & \textbf{8}\\
\hline
\scriptsize{1} & \scriptsize{42.78} & \scriptsize{20.59} & \scriptsize{15.99} & 
\scriptsize{14.61} & \scriptsize{11.05} & \scriptsize{8.89} & \scriptsize{11.56} &
\scriptsize{10.94}\\
\hline
\scriptsize{2} & \scriptsize{99.93} & \scriptsize{52.66} & \scriptsize{38.59} & 
\scriptsize{25.22} & \scriptsize{20.71} & \scriptsize{18.60} & \scriptsize{18.55} &
\scriptsize{18.70}\\
\hline
\scriptsize{3} & \scriptsize{151.03} & \scriptsize{80.49} & \scriptsize{57.33} & 
\scriptsize{39.65} & \scriptsize{28.83} & \scriptsize{36.03} & \scriptsize{44.16} &
\scriptsize{44.76}\\
\hline
\end{tabular} \\ \\

\textbf{2.b} Total execution cpu-times, with data storage:\\

\begin{tabular}{|l|llllllll|}
\hline
\textbf{m} & \textbf{1} & \textbf{2} & \textbf{3} & \textbf{4} & \textbf{5}
& \textbf{6} & \textbf{7} & \textbf{8}\\
\hline
\scriptsize{1} & \scriptsize{49.55} & \scriptsize{28.70} & \scriptsize{26.69} & 
\scriptsize{27.26} & \scriptsize{17.98} & \scriptsize{18.92} & \scriptsize{17.78} &
\scriptsize{17.76}\\
\hline
\scriptsize{2} & \scriptsize{131.66} & \scriptsize{68.50} & \scriptsize{54.33} & 
\scriptsize{40.53} & \scriptsize{38.00} & \scriptsize{38.37} & \scriptsize{36.76} &
\scriptsize{40.69}\\
\hline
\scriptsize{3} & \scriptsize{201.65} & \scriptsize{102.90} & \scriptsize{75.03} & 
\scriptsize{58.94} & \scriptsize{60.97} & \scriptsize{66.52} & \scriptsize{67.92} &
\scriptsize{67.83}\\
\hline
\end{tabular} \\ \\

\section{Analysis of results}
From the results of the previous section, I deduce the following observations:\\
\begin{enumerate}
\item In table \textbf{1.a} the gain in execution speed is good from 1 to 4 processes, 
while from 5 to 8, and in particular in the case m=3, the gain is low; this fact can
be due to the excess load on the dual-processor machine when \textit{nproc} 
\begin{math}>\end{math} 4;
\item In the same table the speedup [5] is quasi-linear from 1 to 4, hence the algorithm
and the parallel environment used are an optimized SPMD implementation if one is
interested only on pure computation time;
\item In table \textbf{2.a} the value \textit{nproc}=4 gives the best performance as
total execution time; hence, if one is interested on the time spent by the master to
verify when the slaves finish their job, the four virtual processors guaranteed by 
Hyper Threading technology show the best efficiency in the case of four running 
processes;
\item With exclusion of the case \textit{nproc}=1, when the master must verify only 
a single process, in the case m=3 the difference between the times rows of \textbf{2.a} 
and \textbf{1.a} is smallest in the case \textit{nproc}=4; hence in this case too the 
total time registered by the master is optimized respect to the slaves execution time;
\item In the case of data storage, the preceding conclusions are not so clear, probably
due to the fact that the reading-writing of small files is in general optimized 
by the RAID 5 technology on a multi-disks system; this fact seems to be confirmed by
the great difference, 50 seconds, from the times registered in the case m=3 and
\textit{nproc}=1 in tables \textbf{.b}; 
\item In table \textbf{2.b} one can note that for m=3 the best result is in the case 
\textit{nproc}=4;
\item In tables \textbf{.b}, in the case m=3 the difference between the times rows is
smallest in the case \textit{nproc}=4 too;
\item From tables \textbf{2.b}, one can note that for m=1 the best speedup (8 processes)
respect to the case \textit{nproc}=1 is 2.78, while for m=3 the best (4 processes) is
3.42; hence the virtual processors seem to have a better performance with a large amount 
of data to compute, probably due to the fact that in this case the parallel computations
have a greater relevance respect to the physical operations on storage system.
\end{enumerate}

\subsection{Conclusions}
From the previous facts one can deduce that a virtual processors technology 
as Hyper Threading can be a good choice for running SPMD programs in the case that
\begin{itemize}
\item the number of parallel processes is equal to the number of virtual processors;
\item the data to be computed have a large amount, particularly when their distribution
among processes and the merging of final results are based on files stored on fast storage
system.
\end{itemize}

\section{Acknowledgements}
I wish to thank Ernesto Montagnoli, Mario Bubola and Filippo Rossetti, chief and members 
of Infrastructures and Systems Area at ICT Department, Riello Group, for the permission 
to execute this work on the ProLiant server and for technical assistence.

\end{document}